\documentclass[floatfix,%
 aip,
 amsmath,amssymb,
 reprint,%
]{revtex4-1}

\usepackage{xcolor, color, soul}
\usepackage{graphicx}
\usepackage{dcolumn}
\usepackage{bm}

\usepackage[utf8]{inputenc}
\usepackage[T1]{fontenc}
\usepackage{mathptmx}
\usepackage{etoolbox}
\makeatletter
\def\@email#1#2{%
 \endgroup
 \patchcmd{\titleblock@produce}
  {\frontmatter@RRAPformat}
  {\frontmatter@RRAPformat{\produce@RRAP{*#1\href{mailto:#2}{#2}}}\frontmatter@RRAPformat}
  {}{}
}%
\makeatother
\usepackage{color,soul}
\begin{document}

\preprint{AIP/123-QED}

\title[]{Measuring Lateral Capillary Forces on Floating Particles using the Moses Effect }
\author{David Shulman}
\affiliation{Department of Chemical Engineering, Ariel University, Ariel, Israel 407000}
\affiliation{Physics Department, Ariel University, Ariel 40700, Israel}
\email{davidshu@ariel.ac.il}

\date{\today}

\begin{abstract}
This study presents a novel and user-friendly technique for detecting the lateral capillary force on a floating spherical particle. The technique leverages the interplay between the capillary attracting forces, hydrostatic pressure forces, and magnetic repulsion forces. A magnetic field is applied to induce a surface curvature in the liquid, resulting in a non-uniform distribution of capillary and hydrostatic pressure forces across the particle's surface. This leads to a stable equilibrium position of the particle at a specific distance from the magnet. The study analyzes the equilibrium position and other relevant parameters in comparison with the developed theory. Classical mechanics and intermolecular forces are applied to establish the theoretical basis for the method, modeling the behavior of the particle in response to the magnetic field, surface curvature, and hydrostatic pressure. The equilibrium position of the particle is determined by numerically solving the balance of forces equation.

\end{abstract}


\maketitle 

\section{Introduction}
The study of the interaction between floating particles and the forces acting upon them has been an active area of research for several decades. These forces are primarily due to interfacial interactions between the particles and the liquid interface, and are responsible for the stability, aggregation, and behavior of particles in the liquid. Several studies have been carried out to calculate these interfacial forces and their impact on the behavior of particles in liquids \cite{nicolson1949interaction,gifford1971attraction,princen1969equilibrium,chan1981interaction,fortes1982attraction,allain1993interaction,danov2010capillary,dixit2012capillary,kralchevsky1995lateral,paunov1993lateral,vassileva2005capillary}.

Nicolson's 1949 paper presented an analytical expression for the capillary interaction force between floating bubbles. To derive this expression, Nicolson first calculated the deformation of the interface caused by a single bubble. Then, he calculated the interaction energy between two bubbles located at a certain distance from each other, assuming that the interface shape for both bubbles was the same. Nicolson also made simplifying assumptions that the bubbles did not tilt and that the contact angle around the contact circle remained zero. These assumptions allowed him to derive an analytical expression for the capillary force \cite{nicolson1949interaction}.\\
Gifford and Scriven \cite{gifford1971attraction}  derived a solution for the capillary attraction between two infinite cylinders, but it was an approximate solution that relied on the assumption that the curvature of the interface near the cylinders was small. Under this assumption, they were able to derive an analytical expression for the interaction force between the cylinders, which is now known as the "Gifford-Scriven logarithmic term". While it is possible to obtain a solution to the complete nonlinear problem in terms of special functions such as elliptic integrals or hypergeometric functions, such solutions are often difficult to obtain and solve numerically. As a result, approximate solutions such as the one derived by Gifford and Scriven are often used in practice as they provide a good approximation to the interaction force in many cases.\\
Another possibility to solve the problem of the capillary interaction between parallel cylinders was demonstrated by Allain and Cloitre  \cite{allain1993interaction}, who used free energy analysis. However, due to the complexity, the resulting equation can only be solved by using numerical methods. Moreover, the Young-Laplace equation linearized in bipolar coordinates can also be used to calculate the attraction force between particles \cite{kralchevsky1992capillary}.\\ Wurger \cite{wurger2006curvature} studied a liquid interface with different principal curvatures and established that the mere presence of a spherical particle led to a deformation field of quadrupolar symmetry; the corresponding “capillary quadrupole moment” was given by the ratio of the particle size and the curvature radius.\\
While theoretical investigations of capillary forces between particles on fluid interfaces are common in the literature, there are relatively few studies where these forces have been investigated experimentally. Some experimental studies have been conducted to investigate the lateral force between two floating particles \cite{dushkin1995lateral,velev1993direct,velev1994capillary}. However, experimental works dedicated to investigating the lateral force on a single floating particle are even more scarce. One possibility to calculate and measure the force on a single spherical particle floating on a curved interface was suggested by Velev et al., who used a piezo-transducer balance \cite{velev1993direct,velev1994capillary}.\\
The researchers investigated a situation where a hydrophobic plate was placed at a specific distance from a floating particle, creating a surface curvature. Due to the interplay between the capillary forces acting on the particle and the gravity and buoyancy forces acting in the opposite direction, the particle reached a stable equilibrium position at a finite distance from the plate. The distance between the particle and the hydrophobic plate was then measured using a microscope.   \\
In the study conducted by Dushkin and colleagues \cite{dushkin1995lateral}, lateral immersion forces were measured systematically using a torsion micro-balance. Specifically, the forces were measured for three different configurations: between two vertical cylinders, between a cylinder and a sphere, and between a sphere and a vertical wall. \\
In this study, we present a new method for measuring the lateral capillary forces on a floating particle based on its response to a surface curvature imposed by a magnetic field. The method is sensitive, convenient, and provides a non-invasive way of measuring the intermolecular forces on the particle. The method is based on the interplay between the capillary attracting forces, the hydrostatic pressure forces, and the magnetic repulsion forces. It is shown that this interplay can lead to the appearance of a stable equilibrium position of the particle at a finite distance from the magnet. The dependence of the equilibrium position of the particle and other parameters is investigated and compared with the developed theory.\\
Overall, this new method represents an important advancement in the field of intermolecular forces and surface chemistry. It has the potential to impact a wide range of applications in various scientific disciplines, including material science, biophysics, and nanotechnology.

\begin{widetext}
\begin{figure*}[!t]
\normalsize
  \begin{minipage}{0.48\textwidth}
     \centering
     \includegraphics[width=1\linewidth]{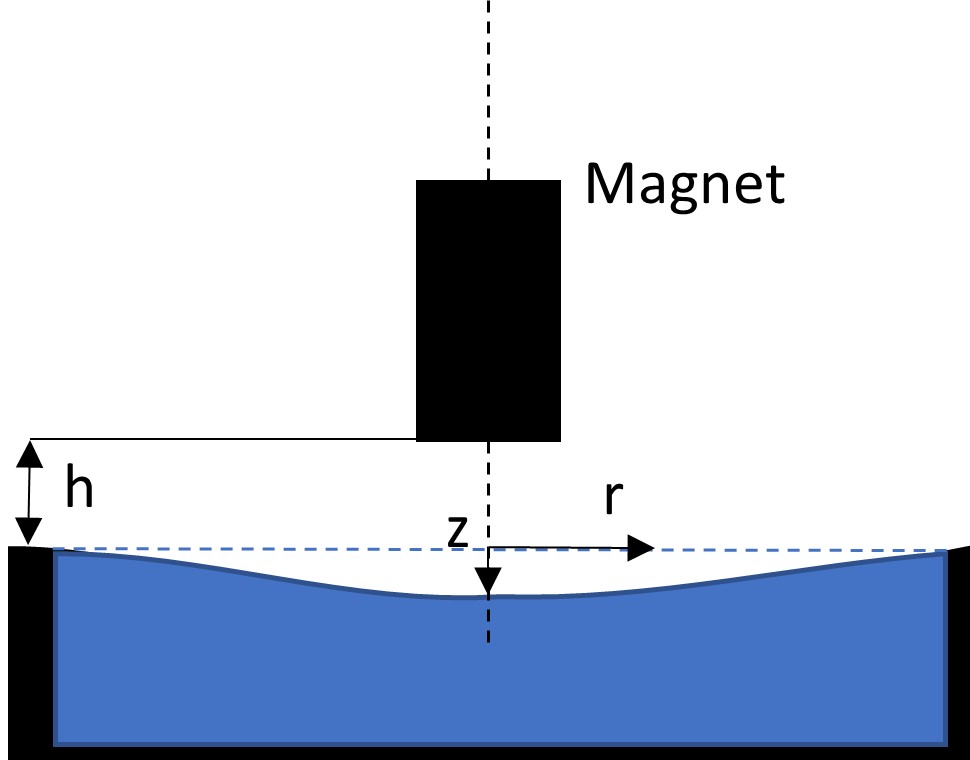}
     \caption{Shape of the near-surface well arising from the deformation of the liquid/vapor interface by the permanent magnetic field  $\vec{B}$ is depicted.}\label{fig:1}
  \end{minipage}\hfill
  \begin{minipage}{0.48\textwidth}
     \centering
     \includegraphics[width=1\linewidth]{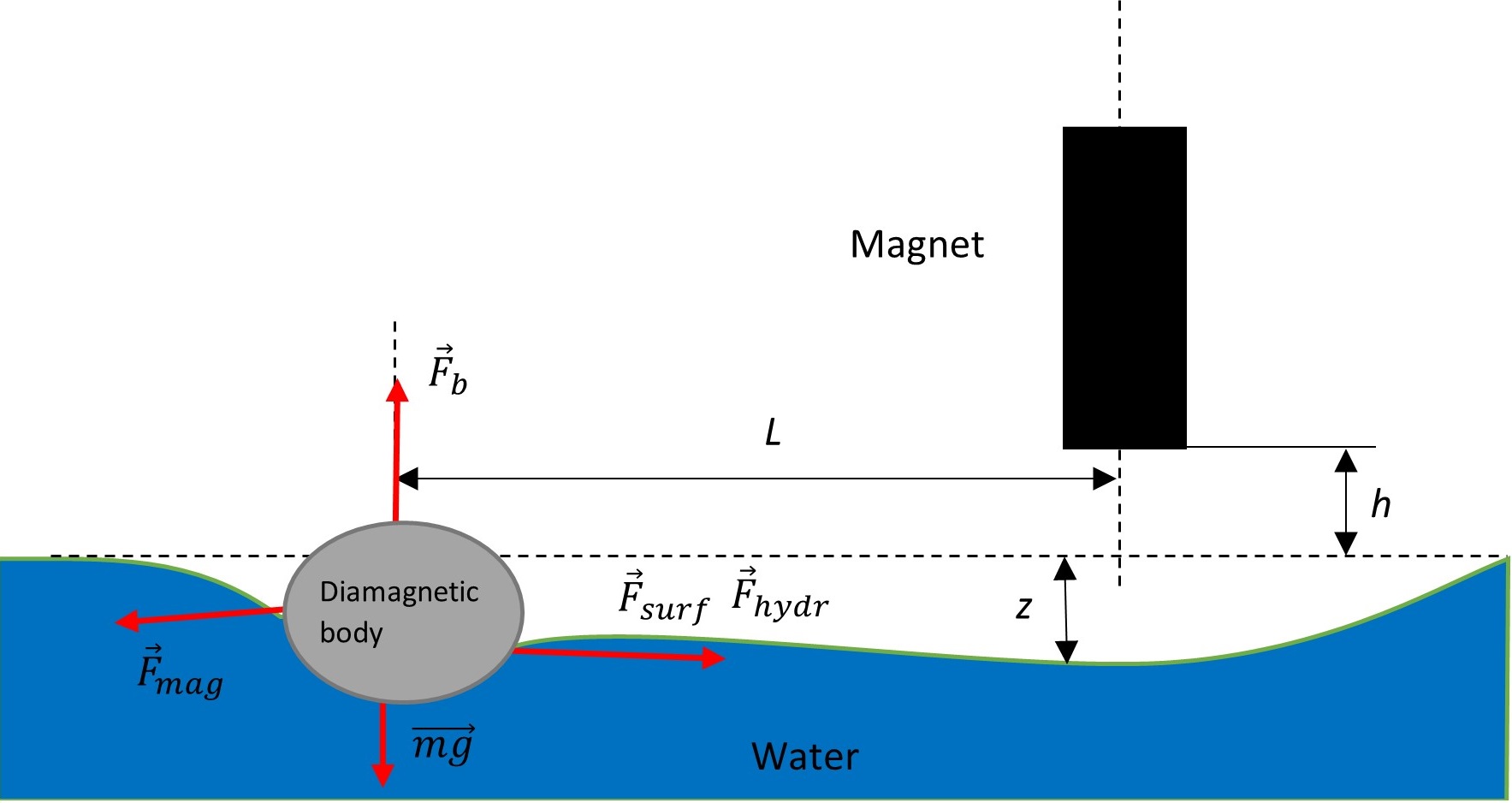}
     \caption{Balance of the forces acting on the floating diamagnetic hydrophobic particle.}\label{fig:2}
  \end{minipage}
\hrulefill
\end{figure*}
\end{widetext} 
\section{Theoretical basis of the suggested experimental method}
\label{section:theor}
\subsection{Introduction}
The capillary force is a type of intermolecular force that arises due to the surface tension of liquids. It is responsible for the tendency of liquids to rise up narrow tubes, for example. The capillary force is also important in determining the behavior of particles in contact with liquid surfaces, and can lead to the aggregation or separation of particles based on their surface properties.

In addition to the capillary forces, the hydrostatic pressure forces on the particle can also be taken into account. The hydrostatic pressure force arises from the pressure difference between the top and bottom surfaces of the particle due to its immersion in a liquid. This force can contribute to the balance of forces acting on the particle in the presence of a magnetic field.

The new method for measuring the lateral capillary force on a particle is based on the interplay between the capillary attracting forces, the hydrostatic pressure forces, and the magnetic repulsion forces due to the particle's diamagnetic properties. The magnetic field is used to impose a surface curvature of the liquid, which results in a non-uniform distribution of capillary and hydrostatic pressure forces along the surface of the particle.

As a result, the particle experiences a net force that causes it to move and ultimately reach a stable equilibrium position at a finite distance from the magnet. The balance of forces acting on the particle on the surface can be expressed mathematically as:

\begin{equation}
\vec{F}_{surf}+\vec{F}_{hydr}-\vec{F}_{magn}=0
\label{balanceF}
\end{equation}

where $\vec{F}_{surf}$ is the capillary force acting on the particle on the surface, $\vec{F}_{hydr}$ is the hydrostatic pressure force acting on the particle, and $\vec{F}_{magn}$ is the magnetic repulsion force acting on the particle.

The theoretical basis for this method involves the application of classical mechanics and intermolecular forces to model the behavior of the particle in response to the magnetic field, surface curvature, and hydrostatic pressure. The dependence of the equilibrium position of the particle and other parameters is investigated and compared with the developed theory to validate the method and provide insights into the nature of intermolecular forces.

Overall, this new method represents an important advancement in the field of intermolecular forces and surface chemistry, and has the potential to impact a wide range of applications in various scientific disciplines.
\subsection{Mathematical formulation}
The classical theory of the Moses effect states that the shape of the "well" formed by the permanent magnet on the diamagnetic liquid surface emerges due to the application of a permanent magnetic field \cite{bormashenko2019moses,myfirstarticle}. For the cylindrically symmetrical magnetic field $\vec{B}\left( r,h \right) $, the following expression predicts the well's shape, displayed in Figure \ref{fig:1}:
\begin{equation}
z\left( r,h\right) =\frac{\chi B^{2}\left( r,h\right) }{2\mu _{0}\rho g}.
\label{wrong_z}
\end{equation}

where $\chi $ and $\rho $ are the magnetic susceptibility and the density of the
liquid, respectively, $\mu_0$ is the magnetic permittivity of vacuum, and $h$ is the separation between the magnet and
non-deformed liquid/vapor interface.

The eq.(\ref{wrong_z}) neglects the surface tension of the liquid/vapor
interface. The Young-Laplace equation defining the shape of the liquid-air interface
deformed by a time-independent magnetic field, which takes into account
surface tension $\gamma $ is derived in \cite{gendelman2019study}:

\begin{equation}
\frac{\partial ^{2}z}{\partial r^{2}}+\frac{1}{r}\frac{\partial z}{\partial r%
}-\frac{\rho g}{\gamma }z=\frac{\chi B^{2}\left( r,h\right) }{2\mu
_{0}\gamma }   \label{z equation}
\end{equation}

with its solution:

\begin{equation}
\begin{split}
z\left( r,h\right) =-\left[ \int_{0}^{r}\frac{\chi B^{2}\left( r^\prime,h\right) }{%
2\mu _{0}\gamma }I_{0}\left( \lambda _{c}^{-1}r^\prime\right) r^\prime dr^\prime\right]
K_{0}\left( \lambda _{c}^{-1}r\right)-\\
-\left[ \int_{r}^{\infty }\frac{\chi
B^{2}\left( r^\prime,h\right) }{2\mu _{0}\gamma }K_{0}\left( \lambda
_{c}^{-1}r^\prime\right) r^\prime dr^\prime\right] I_{0}\left( \lambda _{c}^{-1}r\right)
\label{z}
\end{split}
\end{equation}

Here $I_{0}\left( x\right) $ and $K_{0}\left( x\right) $ are the
modified Bessel functions of the first and the second kind, respectively.

For small curvature $dz/dr\ll 1$, so that $dz/dr\cong \theta $, see Fig. \ref{fig:1}.
The derivative of Eq.(\ref{z}):

\begin{equation}
\begin{split}
\theta \approx \frac{dz}{dr}=\left[ \int_{0}^{r}\frac{\chi B^{2}\left( r^\prime,h\right) }{%
2\mu _{0}\gamma }I_{0}\left( \lambda _{c}^{-1}r^\prime\right) r^\prime dr^\prime\right] \lambda
_{c}^{-1}K_{1}\left( \lambda _{c}^{-1}r\right)-\\ 
-\left[ \int_{r}^{\infty }%
\frac{\chi B^{2}\left( r^\prime,h\right) }{2\mu _{0}\gamma }K_{0}\left( \lambda
_{c}^{-1}r^\prime\right) r^\prime dr^\prime\right] \lambda _{c}^{-1}I_{1}\left( \lambda
_{c}^{-1}r\right)  
\label{theta}
\end{split}
\end{equation}

\begin{widetext} 
\begin{figure*}[!t]
\normalsize
  \begin{minipage}{0.48\textwidth}
     \centering
     \includegraphics[width=1\linewidth]{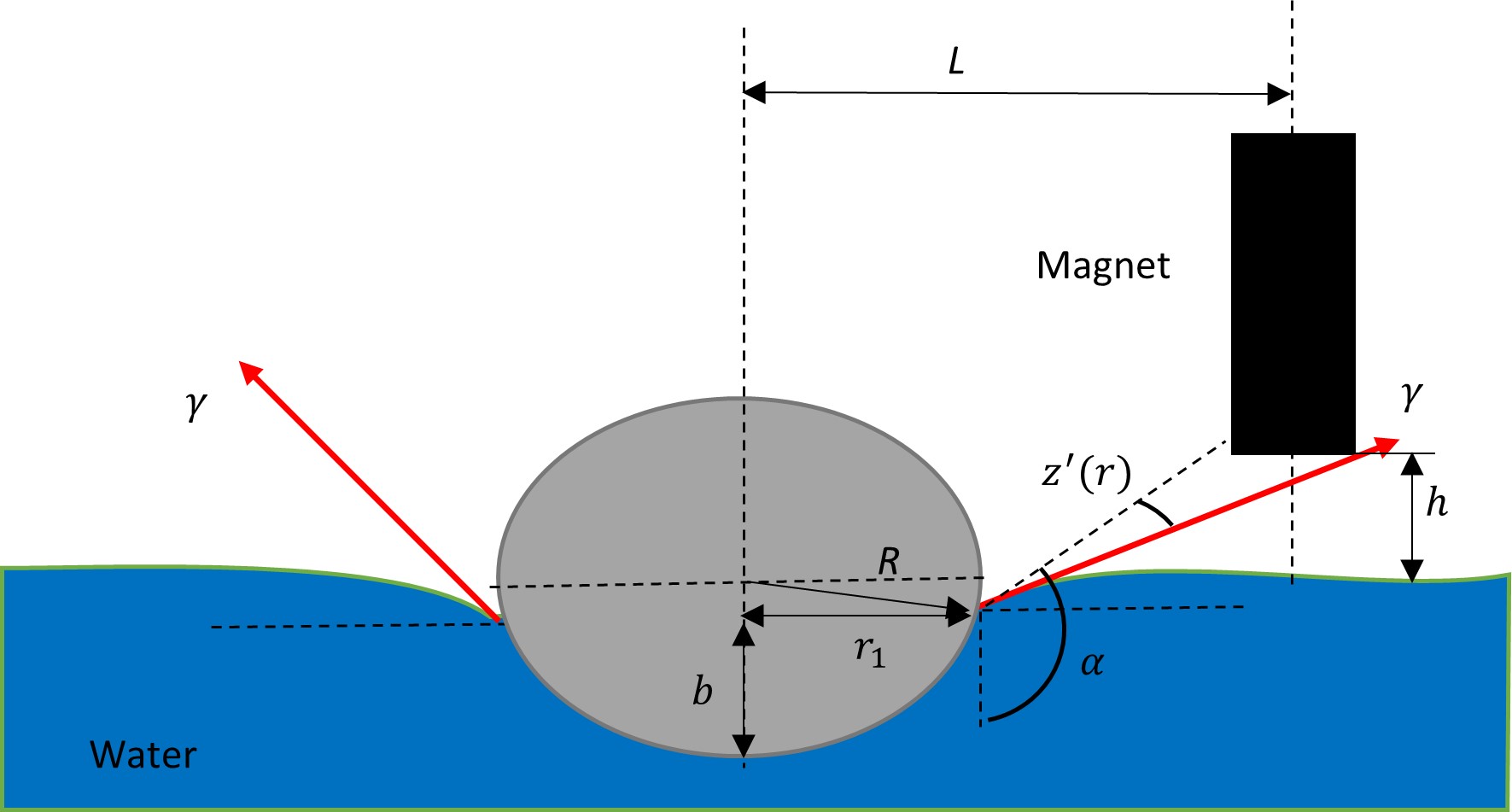}
     \caption{A diamagnetic hydrophobic object floating on the curved surface of water.}\label{fig:3}
  \end{minipage}\hfill
  \begin{minipage}{0.48\textwidth}
     \centering
     \includegraphics[width=1\linewidth]{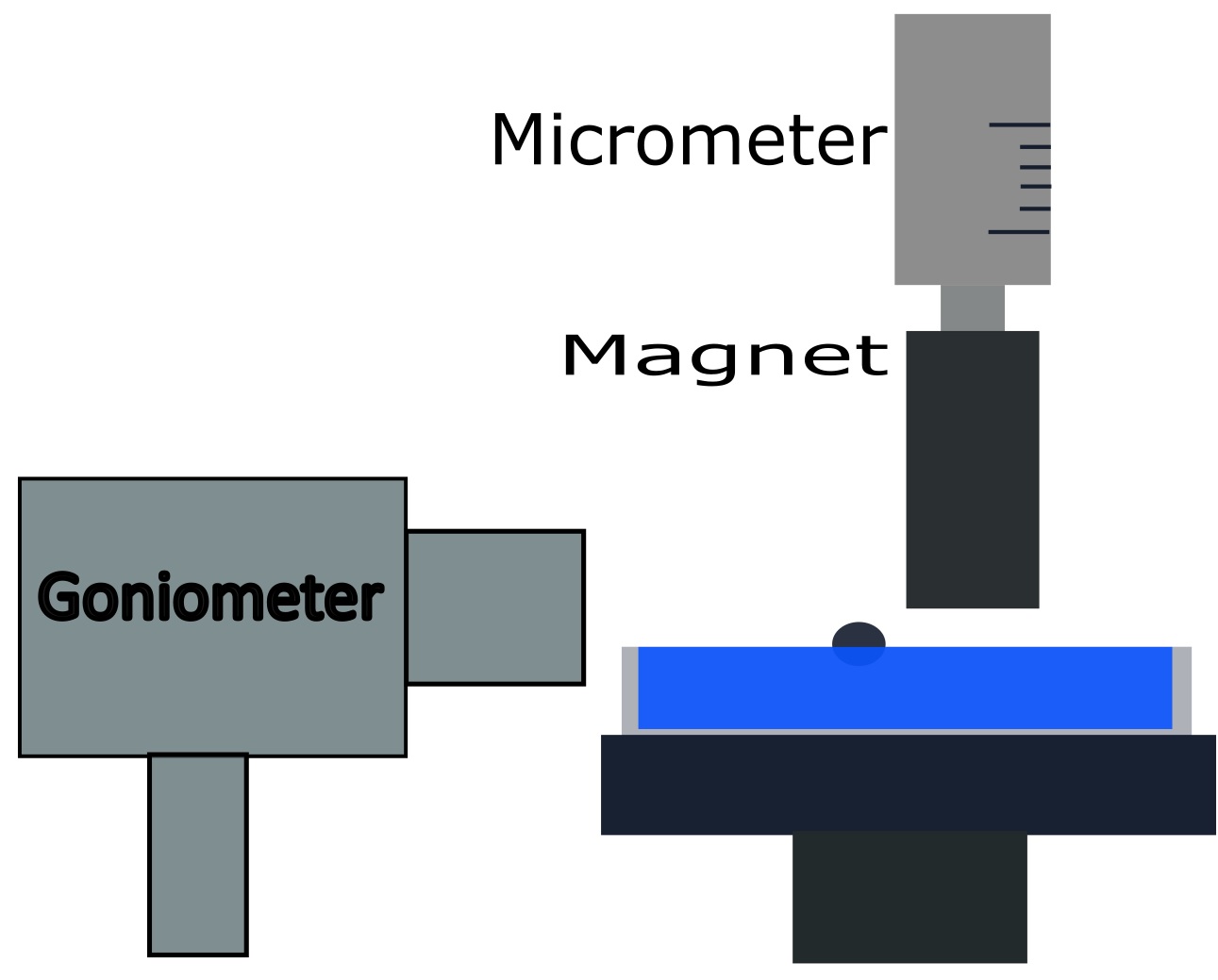}
     \caption{The schematic outline of the experimental setup.}\label{fig:4}
  \end{minipage}
\hrulefill
\end{figure*}
\end{widetext}

The lateral capillary force on the particle can be calculated by integrating the interfacial tension of the meniscus along the contact line \cite{fortes1982attraction}. The lateral projection of the interfacial tension of the meniscus is the component of the tension force that acts in the direction perpendicular to the contact line. It can be calculated using the angle $\alpha$ between the tangent to the liquid surface and the vertical direction, and the slope $z'(r)$ of the disturbed liquid surface caused by the magnetic field. The equation for the lateral projection of the interfacial tension is $\gamma\cos(\alpha-0.5\pi-z^{'}(r))$, see Fig. \ref{fig:3}. The complete integral along the contact line is :

\begin{equation}
F_{surf}(r)=\int_{0}^{2\pi} \gamma R_0 cos(\varphi)cos(\alpha-0.5\pi-z^{'} (r(\varphi)))d\varphi
\label{8}
\end{equation}

where $\gamma$ is the surface tension of the liquid, $R_0$ is the radius of the particle, $\varphi$ is the azimuthal angle, $\alpha$ is the angle between the normal to the surface and the magnetic field, and $z^{'}(r)$ is the first derivative of the surface profile with respect to $r$.\\
To calculate the hydrostatic force on the particle, one can use the following equation:

\begin{equation}
F_{hydr}(r)=-0.5\int_{0}^{2\pi}R_0 cos(\varphi) \bigtriangleup z(r,\varphi)^2 \rho g d \varphi
\end{equation}

In this equation, $\Delta z(r,\varphi)$ represents the height difference between the top and bottom surfaces of the particle at a given position $r$ and azimuthal angle $\varphi$. Other parameters used are the density of the liquid $\rho$, the acceleration due to gravity $g$, and a factor of 0.5 that assumes the particle has a symmetric shape.
The magnetic force on the particle can be calculated as:

\begin{equation}
\begin{split}
\vec{F}_{magn}(r)=-\nabla (U) =\nabla (\vec{m}\cdot \vec{B})=\\=(\nabla \cdot \vec{m}) \vec{B}+ \vec{m}(\nabla \cdot \vec{B}))\approx \vec{m}(\nabla \cdot \vec{B}))=\\ =\frac{V_m \bigtriangleup \chi }{2 \mu_0} \nabla (B \cdot B) 
\label{10}
\end{split}
\end{equation}

where $\vec{m}$ is the magnetic moment of the particle, $\vec{B}$ is the magnetic field, $V_m$ is the volume of the magnetic material in the particle, $\bigtriangleup \chi$ is the difference in magnetic susceptibility between the particle and the surrounding liquid, $\mu_0$ is the permeability of free space, and $\nabla (B \cdot B)$ is the gradient of the square of the magnetic field.
\\
In order to simplify the numerical calculations, it is convenient to convert the volume integral into a surface integral. This can be achieved by applying the divergence theorem, resulting in an expression for the magnetic force that involves the surface integral of the magnetic field squared, weighted by the normal vector and the change in susceptibility. Specifically, the magnetic force can be expressed as:
\begin{equation}
F_{magn}=\frac{\bigtriangleup \chi }{2 \mu_0} \iiint \frac{d(B^2)}{dr}dV =\frac{\bigtriangleup \chi }{2 \mu_0} \oint_{S} B^2 \cdot n dS
\label{11}
\end{equation}

where $\Delta \chi$ represents the relative magnetic susceptibility of the particle, $\mu_0$ is the permeability of free space, $S$ is the surface of the particle, $B$ is the magnetic field, and $\vec{n}$ is the normal vector to the surface.
The corresponding square of the magnetic field $B^2$ was calculated assuming that the magnetic field of a permanent magnet can be approximated as the magnetic field of an ideal solenoid \cite{art_magnet} . For this calculation we use the Python package for the computation of the magnetic fields of \cite{ortner2020magpylib}.

To determine the equilibrium position of the floating particle, the balance of forces equation can be solved numerically using finite element methods or other numerical techniques. The equilibrium position is the point at which the net force on the particle is zero, and it can be used to measure the lateral capillary force acting on the particle. The dependence of the equilibrium position on various parameters such as the surface curvature, magnetic field strength, and particle size can provide insights into the nature of intermolecular forces and their impact on the behavior of particles in liquids.

\section{Experimental arrangement}
The suggested experimental technique is based on determining the coordinates of the stable equilibrium position of the particle on the water surface curved by the magnetic field. A schematic outline of the experimental setup is presented in Fig. \ref{fig:4}. The main elements of the experimental setup, shown in Fig. \ref{fig:4} are:
\begin{enumerate}
\item The sample liquid (deionized water) in the Petri dish with the diameter of 90 $mm$. 
\item A stack of Neodymium permanent magnets.
\item	A special micrometric screw fixing the precise location of the permanent magnet with a precision of 10 $\mu m$. 
\item	A Rame-Hart 500 contact angle goniometer/tensiometer registering the position of the floating particle with an accuracy of 100 $\mu m$. 
\item	The floating polystyrene sphere with the diameter of 2 $mm$.
\end{enumerate}
The petri dish with the diameter of 90 $mm$ was filled with deionized water to a depth of approximately 1 $cm$ and set on top of a frame constructed from a non-ferromagnetic material. A cylindrical neodymium (NdFeB) bar magnet of with the diameter of 22.8 $mm$ and the length of 30 $mm$, supplied by MAGSY, Czech, was used during the measurements. The magnetic field strength at the base center was $B_z\approx 0.6\pm0.01\ T$. 
The magnetic field was measured with a DC/AC GM2 Gauss Meter, manufactured by AlphaLab Inc., USA. For modeling the magnetic field of the magnet an ideal solenoid approximation has been used \cite{art_magnet}. \\
The deionized water was prepared using the synergy UV water purification system from Millipore SAS (France). The specific resistivity of the water used during the experiments was $\rho = 18.2\ M\Omega \times cm\ at\ 25^o\ C$. The deionized water has the following physical properties: density $\rho=998\ kg/m^3$, volume magnetic susceptibility $\chi=-9.04\cdot 10^{-6}$, surface tension $\gamma=72.75\cdot10^{-3}\ N/m$.\\
For the experiments, polystyrene (PE) spheres with the diameter of 2 $mm$, supplied by Cospheric LLC, United States, were used. The polystyrene spheres have the following physical properties \cite{van2009properties,howe1999polymer}: volume magnetic susceptibility $\chi=-8.21\cdot10^{-6}$, density $\rho=1.04\ g/cm^{-3}$  , contact angle $\alpha=100^o\pm 5^o$. The contact angle and the depth of immersion of the PE particles was measured by Rame-Hart 500 contact angle goniometer/tensiometer. \\
The measurements were performed as follows: the magnet was fixed at a certain position on the micrometer precision stage. The equilibrium position of the particle was measured with the goniometer. The micrometer stage lifted the magnet vertically with the step of 0.1 $mm$ and the measurement was repeated. \\
All experiments were carried out at ambient conditions ($P=1\ atm$; $T=25^o\ C$). 
\begin{widetext} 
\begin{figure*}[!t]
\normalsize
  \begin{minipage}{0.48\textwidth}
     \centering
     \includegraphics[width=1\linewidth]{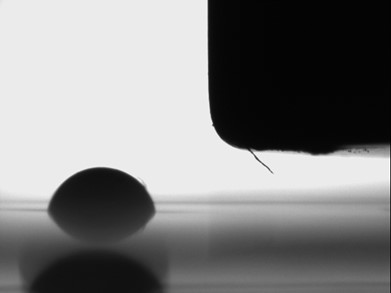}
     \caption{An image captured by the goniometer. A floating particle can be seen on the left, and the magnet can be seen in the top right corner of the image.}\label{fig:5}
  \end{minipage}\hfill
  \begin{minipage}{0.48\textwidth}
     \centering
     \includegraphics[width=1\linewidth]{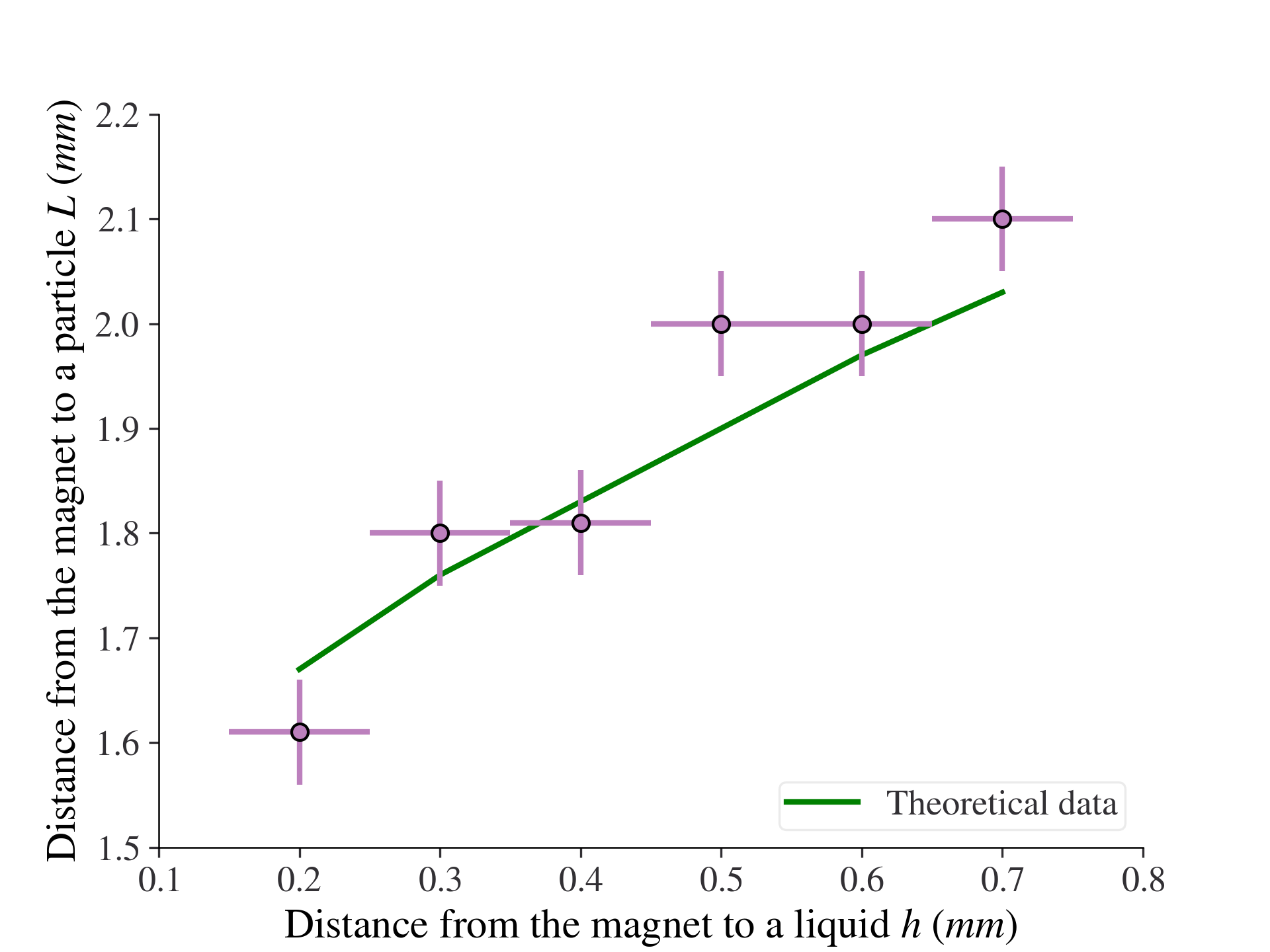}
     \caption{The graph shows the relationship between the distance from the edge of the magnet to the center of the particle, denoted as $L$, and the distance from the edge of the magnet to the liquid surface. The experimental data is represented by dots, while the green line corresponds to the theoretical calculations.}\label{fig:6}
  \end{minipage}
\hrulefill
\end{figure*}
\begin{figure*}[!t]
\normalsize
  \begin{minipage}{0.48\textwidth}
     \centering
     \includegraphics[width=1\linewidth]{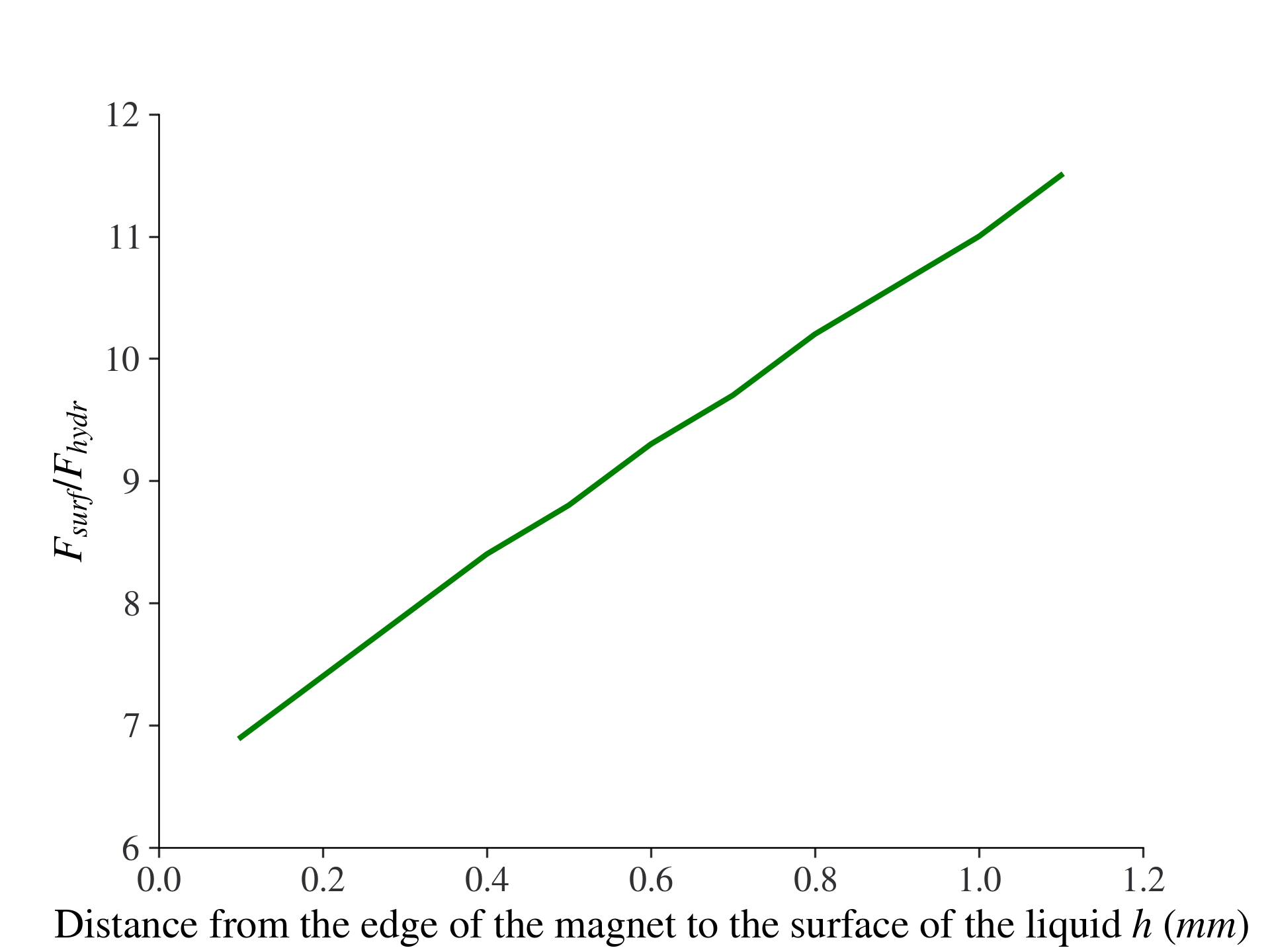}
     \caption{The graph shows the ratio of the magnitude of the surface forces to the magnitude of the forces of hydrostatic pressure for different distances of the magnet to the liquid surface.}\label{fig:7}
  \end{minipage}\hfill
  \begin{minipage}{0.48\textwidth}
     \centering
     \includegraphics[width=1\linewidth]{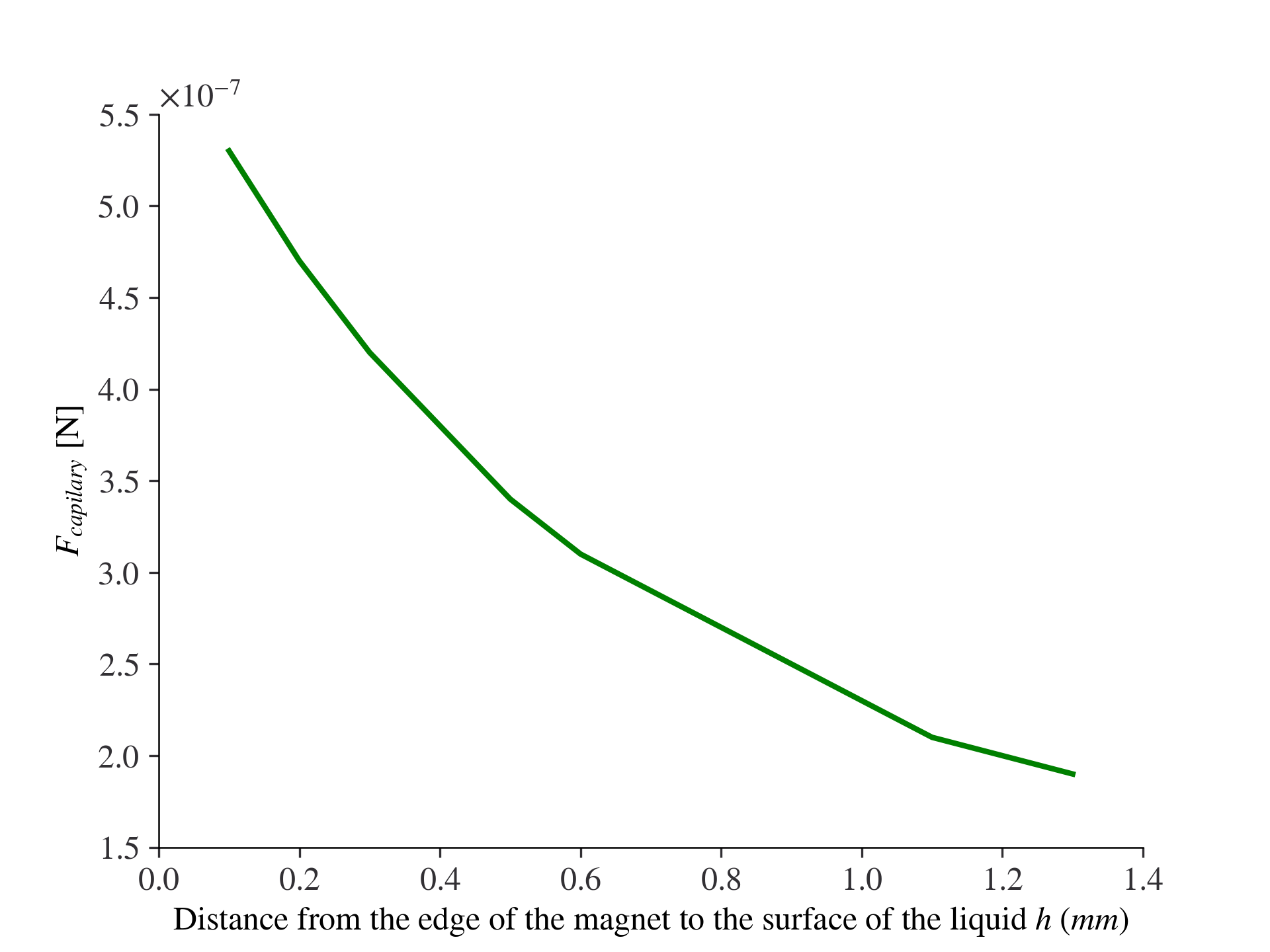}
     \caption{The graph shows the capillary forces $F_{capillary}$,which in our case are equal to the magnetic force on the particle.}\label{fig:8}
  \end{minipage}
\hrulefill
\end{figure*}
\end{widetext} 

\section{Experimental Results and Discussion}

In this study, the capillary meniscus interaction between a floating particle and a surface deformation was investigated, and a setup for measuring the capillary forces using the Moses effect was developed and employed for the experimental procedures. Fig. \ref{fig:5} shows an image captured by the goniometer, where a meniscus formed around the floating particle is clearly visible, while the well formed by the magnetic field cannot be seen due to its small size.

The theoretical calculations based on the numerical solution of equation (\ref{balanceF}), as described in Section 2, are illustrated with the green line in Fig. \ref{fig:6}. The graph demonstrates that the horizontal distance between the center of the floating particle and the edge of the magnet increases with the increase of the vertical distance of the magnet from the liquid. This effect can be explained by considering the change in the forces acting on the particle due to the magnet movement. As the magnet moves vertically, the horizontal component of the magnetic field acting on the particle does not change as much as the capillary attraction forces. The capillary attraction forces, however, decrease significantly since the well and, therefore, the surface slope decrease when the vertical distance from the magnet increases. The experimental error bars are also demonstrated, and it is evident that the theoretical curve falls well within the error bars, thus confirming the theoretical model by the experimental investigations.

Fig. \ref{fig:7} shows the ratio of the surface forces to the forces of hydrostatic pressure in the equilibrium position. This ratio almost linearly increases with the increase of the distance from the magnet to the liquid. The near-linear dependence can be explained by the fact that hydrostatic pressure acts on the surface, meaning that it has a second-degree dependence on the surface area, while surface tension acts on the contact line, having a first-degree dependence on the area. Therefore, as the curvature becomes less pronounced, we get a linear dependence between the forces of hydrostatic pressure and surface forces. From this, we can also conclude that the surface forces are about 10 times bigger than the hydrostatic forces, and in most cases, only the surface forces need to be taken into account.

Fig. \ref{fig:8} shows the sum of all capillary forces (surface forces and the forces of hydrostatic pressure), which is equal to the magnetic force as described by equation (\ref{balanceF}). The graph demonstrates that the capillary forces decrease exponentially as the distance of the magnet from the liquid surface increases. This decrease in the capillary forces can be explained by the decreasing surface slope caused by the magnetic field.

Although the results of this paper apply exclusively to spherical particles partially immersed in fluid, the developed theoretical model can be applied without substantial transformation to other geometries of interest and to superhydrophobic bodies not immersed in a liquid.

\section{Conclusion}

In this study, we have presented a method for measuring the capillary meniscus interaction between a floating particle and a surface deformation. The method is based on the Moses effect, and it was shown to produce reliable results that agree well with theoretical calculations.
\\
The experimental results confirmed that the horizontal distance between the center of the floating particle and the edge of the magnet increases with the increase of the vertical distance of the magnet from the liquid. The near-linear dependence of the ratio of the surface forces to the forces of hydrostatic pressure on the distance from the magnet to the liquid was explained by the different dependence of these forces on the surface area.
\\
The developed theoretical model can be applied without substantial transformation to other geometries of interest and to superhydrophobic bodies non-immersed in a liquid. We expect that the method and the theoretical model will be useful for a range of applications, including the design of micro- and nanoscale devices that rely on capillary forces.\\
Previous research has shown that the contact angle, which is an important parameter in measuring lateral capillary forces, can vary with time depending on factors such as vapor environment and temperature \cite{xinping2005dynamic,yekta1988influences,kwok1999contact}. The contact method for measuring these forces requires a high level of cleanliness for the contact element, and the element must be cleaned after each experiment. Additionally, the contact angle of the element can change during the experiment, making non-contact measurement methods advantageous. The experimental setup described in this study can be utilized for a variety of applications, such as measuring capillary angles, fluid properties, and particle surface properties.\\
Overall, our study highlights the importance of understanding and characterizing capillary forces, and we hope that it will inspire further research in this area.

\subsection*{Conflict of Interest}
The author has no conflicts to disclose.





\bibliography{mybibfile}

\begin{thebibliography}{26}%
\makeatletter
\providecommand \@ifxundefined [1]{%
 \@ifx{#1\undefined}
}%
\providecommand \@ifnum [1]{%
 \ifnum #1\expandafter \@firstoftwo
 \else \expandafter \@secondoftwo
 \fi
}%
\providecommand \@ifx [1]{%
 \ifx #1\expandafter \@firstoftwo
 \else \expandafter \@secondoftwo
 \fi
}%
\providecommand \natexlab [1]{#1}%
\providecommand \enquote  [1]{``#1''}%
\providecommand \bibnamefont  [1]{#1}%
\providecommand \bibfnamefont [1]{#1}%
\providecommand \citenamefont [1]{#1}%
\providecommand \href@noop [0]{\@secondoftwo}%
\providecommand \href [0]{\begingroup \@sanitize@url \@href}%
\providecommand \@href[1]{\@@startlink{#1}\@@href}%
\providecommand \@@href[1]{\endgroup#1\@@endlink}%
\providecommand \@sanitize@url [0]{\catcode `\\12\catcode `\$12\catcode
  `\&12\catcode `\#12\catcode `\^12\catcode `\_12\catcode `\%12\relax}%
\providecommand \@@startlink[1]{}%
\providecommand \@@endlink[0]{}%
\providecommand \url  [0]{\begingroup\@sanitize@url \@url }%
\providecommand \@url [1]{\endgroup\@href {#1}{\urlprefix }}%
\providecommand \urlprefix  [0]{URL }%
\providecommand \Eprint [0]{\href }%
\providecommand \doibase [0]{https://doi.org/}%
\providecommand \selectlanguage [0]{\@gobble}%
\providecommand \bibinfo  [0]{\@secondoftwo}%
\providecommand \bibfield  [0]{\@secondoftwo}%
\providecommand \translation [1]{[#1]}%
\providecommand \BibitemOpen [0]{}%
\providecommand \bibitemStop [0]{}%
\providecommand \bibitemNoStop [0]{.\EOS\space}%
\providecommand \EOS [0]{\spacefactor3000\relax}%
\providecommand \BibitemShut  [1]{\csname bibitem#1\endcsname}%
\let\auto@bib@innerbib\@empty
\bibitem [{\citenamefont {Nicolson}(1949)}]{nicolson1949interaction}%
  \BibitemOpen
  \bibfield  {author} {\bibinfo {author} {\bibfnamefont {M.}~\bibnamefont
  {Nicolson}},\ }\bibfield  {booktitle} {\emph {\bibinfo {booktitle}
  {Mathematical Proceedings of the Cambridge Philosophical Society}},\
  }\href@noop {} {\bibfield  {journal} {\bibinfo  {journal} {Cambridge
  University Press}\ }\textbf {\bibinfo {volume} {45}},\ \bibinfo {pages} {288}
  (\bibinfo {year} {1949})}\BibitemShut {NoStop}%
\bibitem [{\citenamefont {Gifford}\ and\ \citenamefont
  {Scriven}(1971)}]{gifford1971attraction}%
  \BibitemOpen
  \bibfield  {author} {\bibinfo {author} {\bibfnamefont {W.}~\bibnamefont
  {Gifford}}\ and\ \bibinfo {author} {\bibfnamefont {L.}~\bibnamefont
  {Scriven}},\ }\href@noop {} {\bibfield  {journal} {\bibinfo  {journal}
  {Chemical Engineering Science}\ }\textbf {\bibinfo {volume} {26}},\ \bibinfo
  {pages} {287} (\bibinfo {year} {1971})}\BibitemShut {NoStop}%
\bibitem [{\citenamefont {Princen}(1969)}]{princen1969equilibrium}%
  \BibitemOpen
  \bibfield  {author} {\bibinfo {author} {\bibfnamefont {H.}~\bibnamefont
  {Princen}},\ }\href@noop {} {\bibfield  {journal} {\bibinfo  {journal}
  {Surface and colloid science}\ }\textbf {\bibinfo {volume} {2}},\ \bibinfo
  {pages} {1} (\bibinfo {year} {1969})}\BibitemShut {NoStop}%
\bibitem [{\citenamefont {Chan}\ \emph {et~al.}(1981)\citenamefont {Chan},
  \citenamefont {Henry~Jr},\ and\ \citenamefont {White}}]{chan1981interaction}%
  \BibitemOpen
  \bibfield  {author} {\bibinfo {author} {\bibfnamefont {D.}~\bibnamefont
  {Chan}}, \bibinfo {author} {\bibfnamefont {J.}~\bibnamefont {Henry~Jr}},\
  and\ \bibinfo {author} {\bibfnamefont {L.}~\bibnamefont {White}},\
  }\href@noop {} {\bibfield  {journal} {\bibinfo  {journal} {Journal of Colloid
  and Interface Science}\ }\textbf {\bibinfo {volume} {79}},\ \bibinfo {pages}
  {410} (\bibinfo {year} {1981})}\BibitemShut {NoStop}%
\bibitem [{\citenamefont {Fortes}(1982)}]{fortes1982attraction}%
  \BibitemOpen
  \bibfield  {author} {\bibinfo {author} {\bibfnamefont {M.}~\bibnamefont
  {Fortes}},\ }\href@noop {} {\bibfield  {journal} {\bibinfo  {journal}
  {Canadian Journal of Chemistry}\ }\textbf {\bibinfo {volume} {60}},\ \bibinfo
  {pages} {2889} (\bibinfo {year} {1982})}\BibitemShut {NoStop}%
\bibitem [{\citenamefont {Allain}\ and\ \citenamefont
  {Cloitre}(1993)}]{allain1993interaction}%
  \BibitemOpen
  \bibfield  {author} {\bibinfo {author} {\bibfnamefont {C.}~\bibnamefont
  {Allain}}\ and\ \bibinfo {author} {\bibfnamefont {M.}~\bibnamefont
  {Cloitre}},\ }\href@noop {} {\bibfield  {journal} {\bibinfo  {journal}
  {Journal of colloid and interface science}\ }\textbf {\bibinfo {volume}
  {157}},\ \bibinfo {pages} {269} (\bibinfo {year} {1993})}\BibitemShut
  {NoStop}%
\bibitem [{\citenamefont {Danov}\ and\ \citenamefont
  {Kralchevsky}(2010)}]{danov2010capillary}%
  \BibitemOpen
  \bibfield  {author} {\bibinfo {author} {\bibfnamefont {K.~D.}\ \bibnamefont
  {Danov}}\ and\ \bibinfo {author} {\bibfnamefont {P.~A.}\ \bibnamefont
  {Kralchevsky}},\ }\href@noop {} {\bibfield  {journal} {\bibinfo  {journal}
  {Advances in colloid and interface science}\ }\textbf {\bibinfo {volume}
  {154}},\ \bibinfo {pages} {91} (\bibinfo {year} {2010})}\BibitemShut
  {NoStop}%
\bibitem [{\citenamefont {Dixit}\ and\ \citenamefont
  {Homsy}(2012)}]{dixit2012capillary}%
  \BibitemOpen
  \bibfield  {author} {\bibinfo {author} {\bibfnamefont {H.~N.}\ \bibnamefont
  {Dixit}}\ and\ \bibinfo {author} {\bibfnamefont {G.}~\bibnamefont {Homsy}},\
  }\href@noop {} {\bibfield  {journal} {\bibinfo  {journal} {Physics of
  Fluids}\ }\textbf {\bibinfo {volume} {24}},\ \bibinfo {pages} {122102}
  (\bibinfo {year} {2012})}\BibitemShut {NoStop}%
\bibitem [{\citenamefont {Kralchevsky}\ \emph {et~al.}(1995)\citenamefont
  {Kralchevsky}, \citenamefont {Paunov},\ and\ \citenamefont
  {Nagayama}}]{kralchevsky1995lateral}%
  \BibitemOpen
  \bibfield  {author} {\bibinfo {author} {\bibfnamefont {P.}~\bibnamefont
  {Kralchevsky}}, \bibinfo {author} {\bibfnamefont {V.}~\bibnamefont
  {Paunov}},\ and\ \bibinfo {author} {\bibfnamefont {K.}~\bibnamefont
  {Nagayama}},\ }\href@noop {} {\bibfield  {journal} {\bibinfo  {journal}
  {Journal of Fluid Mechanics}\ }\textbf {\bibinfo {volume} {299}},\ \bibinfo
  {pages} {105} (\bibinfo {year} {1995})}\BibitemShut {NoStop}%
\bibitem [{\citenamefont {Paunov}\ \emph {et~al.}(1993)\citenamefont {Paunov},
  \citenamefont {Kralchevsky}, \citenamefont {Denkov},\ and\ \citenamefont
  {Nagayama}}]{paunov1993lateral}%
  \BibitemOpen
  \bibfield  {author} {\bibinfo {author} {\bibfnamefont {V.}~\bibnamefont
  {Paunov}}, \bibinfo {author} {\bibfnamefont {P.}~\bibnamefont {Kralchevsky}},
  \bibinfo {author} {\bibfnamefont {N.}~\bibnamefont {Denkov}},\ and\ \bibinfo
  {author} {\bibfnamefont {K.}~\bibnamefont {Nagayama}},\ }\href@noop {}
  {\bibfield  {journal} {\bibinfo  {journal} {Journal of colloid and interface
  science}\ }\textbf {\bibinfo {volume} {157}},\ \bibinfo {pages} {100}
  (\bibinfo {year} {1993})}\BibitemShut {NoStop}%
\bibitem [{\citenamefont {Vassileva}\ \emph {et~al.}(2005)\citenamefont
  {Vassileva}, \citenamefont {van~den Ende}, \citenamefont {Mugele},\ and\
  \citenamefont {Mellema}}]{vassileva2005capillary}%
  \BibitemOpen
  \bibfield  {author} {\bibinfo {author} {\bibfnamefont {N.~D.}\ \bibnamefont
  {Vassileva}}, \bibinfo {author} {\bibfnamefont {D.}~\bibnamefont {van~den
  Ende}}, \bibinfo {author} {\bibfnamefont {F.}~\bibnamefont {Mugele}},\ and\
  \bibinfo {author} {\bibfnamefont {J.}~\bibnamefont {Mellema}},\ }\href@noop
  {} {\bibfield  {journal} {\bibinfo  {journal} {Langmuir}\ }\textbf {\bibinfo
  {volume} {21}},\ \bibinfo {pages} {11190} (\bibinfo {year}
  {2005})}\BibitemShut {NoStop}%
\bibitem [{\citenamefont {Kralchevsky}\ \emph {et~al.}(1992)\citenamefont
  {Kralchevsky}, \citenamefont {Paunov}, \citenamefont {Ivanov},\ and\
  \citenamefont {Nagayama}}]{kralchevsky1992capillary}%
  \BibitemOpen
  \bibfield  {author} {\bibinfo {author} {\bibfnamefont {P.}~\bibnamefont
  {Kralchevsky}}, \bibinfo {author} {\bibfnamefont {V.}~\bibnamefont {Paunov}},
  \bibinfo {author} {\bibfnamefont {I.}~\bibnamefont {Ivanov}},\ and\ \bibinfo
  {author} {\bibfnamefont {K.}~\bibnamefont {Nagayama}},\ }\href@noop {}
  {\bibfield  {journal} {\bibinfo  {journal} {Journal of Colloid and Interface
  Science}\ }\textbf {\bibinfo {volume} {151}},\ \bibinfo {pages} {79}
  (\bibinfo {year} {1992})}\BibitemShut {NoStop}%
\bibitem [{\citenamefont {W{\"u}rger}(2006)}]{wurger2006curvature}%
  \BibitemOpen
  \bibfield  {author} {\bibinfo {author} {\bibfnamefont {A.}~\bibnamefont
  {W{\"u}rger}},\ }\href@noop {} {\bibfield  {journal} {\bibinfo  {journal}
  {Physical Review E}\ }\textbf {\bibinfo {volume} {74}},\ \bibinfo {pages}
  {041402} (\bibinfo {year} {2006})}\BibitemShut {NoStop}%
\bibitem [{\citenamefont {Dushkin}\ \emph {et~al.}(1995)\citenamefont
  {Dushkin}, \citenamefont {Kralchevsky}, \citenamefont {Yoshimura},\ and\
  \citenamefont {Nagayama}}]{dushkin1995lateral}%
  \BibitemOpen
  \bibfield  {author} {\bibinfo {author} {\bibfnamefont {C.}~\bibnamefont
  {Dushkin}}, \bibinfo {author} {\bibfnamefont {P.}~\bibnamefont
  {Kralchevsky}}, \bibinfo {author} {\bibfnamefont {H.}~\bibnamefont
  {Yoshimura}},\ and\ \bibinfo {author} {\bibfnamefont {K.}~\bibnamefont
  {Nagayama}},\ }\href@noop {} {\bibfield  {journal} {\bibinfo  {journal}
  {Physical review letters}\ }\textbf {\bibinfo {volume} {75}},\ \bibinfo
  {pages} {3454} (\bibinfo {year} {1995})}\BibitemShut {NoStop}%
\bibitem [{\citenamefont {Velev}\ \emph {et~al.}(1993)\citenamefont {Velev},
  \citenamefont {Denkov}, \citenamefont {Paunov}, \citenamefont {Kralchevsky},\
  and\ \citenamefont {Nagayama}}]{velev1993direct}%
  \BibitemOpen
  \bibfield  {author} {\bibinfo {author} {\bibfnamefont {O.~D.}\ \bibnamefont
  {Velev}}, \bibinfo {author} {\bibfnamefont {N.~D.}\ \bibnamefont {Denkov}},
  \bibinfo {author} {\bibfnamefont {V.~N.}\ \bibnamefont {Paunov}}, \bibinfo
  {author} {\bibfnamefont {P.~A.}\ \bibnamefont {Kralchevsky}},\ and\ \bibinfo
  {author} {\bibfnamefont {K.}~\bibnamefont {Nagayama}},\ }\href@noop {}
  {\bibfield  {journal} {\bibinfo  {journal} {Langmuir}\ }\textbf {\bibinfo
  {volume} {9}},\ \bibinfo {pages} {3702} (\bibinfo {year} {1993})}\BibitemShut
  {NoStop}%
\bibitem [{\citenamefont {Velev}\ \emph {et~al.}(1994)\citenamefont {Velev},
  \citenamefont {Denkov}, \citenamefont {Paunov}, \citenamefont {Kralchevsky},\
  and\ \citenamefont {Nagayama}}]{velev1994capillary}%
  \BibitemOpen
  \bibfield  {author} {\bibinfo {author} {\bibfnamefont {O.~D.}\ \bibnamefont
  {Velev}}, \bibinfo {author} {\bibfnamefont {N.~D.}\ \bibnamefont {Denkov}},
  \bibinfo {author} {\bibfnamefont {V.~N.}\ \bibnamefont {Paunov}}, \bibinfo
  {author} {\bibfnamefont {P.~A.}\ \bibnamefont {Kralchevsky}},\ and\ \bibinfo
  {author} {\bibfnamefont {K.}~\bibnamefont {Nagayama}},\ }\href@noop {}
  {\bibfield  {journal} {\bibinfo  {journal} {Journal of colloid and interface
  science}\ }\textbf {\bibinfo {volume} {167}},\ \bibinfo {pages} {66}
  (\bibinfo {year} {1994})}\BibitemShut {NoStop}%
\bibitem [{\citenamefont {Bormashenko}(2019)}]{bormashenko2019moses}%
  \BibitemOpen
  \bibfield  {author} {\bibinfo {author} {\bibfnamefont {E.}~\bibnamefont
  {Bormashenko}},\ }\href@noop {} {\bibfield  {journal} {\bibinfo  {journal}
  {Advances in colloid and interface science}\ }\textbf {\bibinfo {volume}
  {269}},\ \bibinfo {pages} {1} (\bibinfo {year} {2019})}\BibitemShut {NoStop}%
\bibitem [{\citenamefont {Shulman}\ \emph {et~al.}(2023)\citenamefont
  {Shulman}, \citenamefont {Lewkowicz},\ and\ \citenamefont
  {Bormashenko}}]{myfirstarticle}%
  \BibitemOpen
  \bibfield  {author} {\bibinfo {author} {\bibfnamefont {D.}~\bibnamefont
  {Shulman}}, \bibinfo {author} {\bibfnamefont {M.}~\bibnamefont {Lewkowicz}},\
  and\ \bibinfo {author} {\bibfnamefont {E.}~\bibnamefont {Bormashenko}},\
  }\href {https://doi.org/https://doi.org/10.1016/j.jmmm.2023.170553}
  {\bibfield  {journal} {\bibinfo  {journal} {Journal of Magnetism and Magnetic
  Materials}\ ,\ \bibinfo {pages} {170553}} (\bibinfo {year}
  {2023})}\BibitemShut {NoStop}%
\bibitem [{\citenamefont {Gendelman}\ \emph {et~al.}(2019)\citenamefont
  {Gendelman}, \citenamefont {Frenkel}, \citenamefont {Fliagin}, \citenamefont
  {Ivanova}, \citenamefont {Danchuk}, \citenamefont {Legchenkova},
  \citenamefont {Vilk},\ and\ \citenamefont
  {Bormashenko}}]{gendelman2019study}%
  \BibitemOpen
  \bibfield  {author} {\bibinfo {author} {\bibfnamefont {O.}~\bibnamefont
  {Gendelman}}, \bibinfo {author} {\bibfnamefont {M.}~\bibnamefont {Frenkel}},
  \bibinfo {author} {\bibfnamefont {V.}~\bibnamefont {Fliagin}}, \bibinfo
  {author} {\bibfnamefont {N.}~\bibnamefont {Ivanova}}, \bibinfo {author}
  {\bibfnamefont {V.}~\bibnamefont {Danchuk}}, \bibinfo {author} {\bibfnamefont
  {I.}~\bibnamefont {Legchenkova}}, \bibinfo {author} {\bibfnamefont
  {A.}~\bibnamefont {Vilk}},\ and\ \bibinfo {author} {\bibfnamefont
  {E.}~\bibnamefont {Bormashenko}},\ }\href@noop {} {\bibfield  {journal}
  {\bibinfo  {journal} {Surface Innovations}\ }\textbf {\bibinfo {volume}
  {7}},\ \bibinfo {pages} {194} (\bibinfo {year} {2019})}\BibitemShut {NoStop}%
\bibitem [{\citenamefont {Shulman}(2023)}]{art_magnet}%
  \BibitemOpen
  \bibfield  {author} {\bibinfo {author} {\bibfnamefont {D.}~\bibnamefont
  {Shulman}},\ }\href@noop {} {\bibfield  {journal} {\bibinfo  {journal}
  {Journal of Magnetics}\ }\textbf {\bibinfo {volume} {28}},\ \bibinfo {pages}
  {1} (\bibinfo {year} {2023})}\BibitemShut {NoStop}%
\bibitem [{\citenamefont {Ortner}\ and\ \citenamefont
  {Bandeira}(2020)}]{ortner2020magpylib}%
  \BibitemOpen
  \bibfield  {author} {\bibinfo {author} {\bibfnamefont {M.}~\bibnamefont
  {Ortner}}\ and\ \bibinfo {author} {\bibfnamefont {L.~G.~C.}\ \bibnamefont
  {Bandeira}},\ }\href@noop {} {\bibfield  {journal} {\bibinfo  {journal}
  {SoftwareX}\ }\textbf {\bibinfo {volume} {11}},\ \bibinfo {pages} {100466}
  (\bibinfo {year} {2020})}\BibitemShut {NoStop}%
\bibitem [{\citenamefont {Van~Krevelen}\ and\ \citenamefont
  {Te~Nijenhuis}(2009)}]{van2009properties}%
  \BibitemOpen
  \bibfield  {author} {\bibinfo {author} {\bibfnamefont {D.~W.}\ \bibnamefont
  {Van~Krevelen}}\ and\ \bibinfo {author} {\bibfnamefont {K.}~\bibnamefont
  {Te~Nijenhuis}},\ }\href@noop {} {\emph {\bibinfo {title} {Properties of
  polymers: their correlation with chemical structure; their numerical
  estimation and prediction from additive group contributions}}}\ (\bibinfo
  {publisher} {Elsevier},\ \bibinfo {year} {2009})\BibitemShut {NoStop}%
\bibitem [{\citenamefont {Howe}\ and\ \citenamefont
  {Mark}(1999)}]{howe1999polymer}%
  \BibitemOpen
  \bibfield  {author} {\bibinfo {author} {\bibfnamefont {D.}~\bibnamefont
  {Howe}}\ and\ \bibinfo {author} {\bibfnamefont {J.}~\bibnamefont {Mark}},\
  }\href@noop {} {\bibinfo {title} {Polymer data handbook}} (\bibinfo {year}
  {1999})\BibitemShut {NoStop}%
\bibitem [{\citenamefont {Xinping}\ \emph {et~al.}(2005)\citenamefont
  {Xinping}, \citenamefont {Zhifang},\ and\ \citenamefont
  {Zhiquan}}]{xinping2005dynamic}%
  \BibitemOpen
  \bibfield  {author} {\bibinfo {author} {\bibfnamefont {W.}~\bibnamefont
  {Xinping}}, \bibinfo {author} {\bibfnamefont {C.}~\bibnamefont {Zhifang}},\
  and\ \bibinfo {author} {\bibfnamefont {S.}~\bibnamefont {Zhiquan}},\
  }\href@noop {} {\bibfield  {journal} {\bibinfo  {journal} {Science in China
  Series B-Chemistry}\ }\textbf {\bibinfo {volume} {48}},\ \bibinfo {pages}
  {553} (\bibinfo {year} {2005})}\BibitemShut {NoStop}%
\bibitem [{\citenamefont {Yekta-Fard}\ and\ \citenamefont
  {Ponter}(1988)}]{yekta1988influences}%
  \BibitemOpen
  \bibfield  {author} {\bibinfo {author} {\bibfnamefont {M.}~\bibnamefont
  {Yekta-Fard}}\ and\ \bibinfo {author} {\bibfnamefont {A.~B.}\ \bibnamefont
  {Ponter}},\ }\href@noop {} {\bibfield  {journal} {\bibinfo  {journal}
  {Journal of colloid and interface science}\ }\textbf {\bibinfo {volume}
  {126}},\ \bibinfo {pages} {134} (\bibinfo {year} {1988})}\BibitemShut
  {NoStop}%
\bibitem [{\citenamefont {Kwok}\ and\ \citenamefont
  {Neumann}(1999)}]{kwok1999contact}%
  \BibitemOpen
  \bibfield  {author} {\bibinfo {author} {\bibfnamefont {D.~Y.}\ \bibnamefont
  {Kwok}}\ and\ \bibinfo {author} {\bibfnamefont {A.~W.}\ \bibnamefont
  {Neumann}},\ }\href@noop {} {\bibfield  {journal} {\bibinfo  {journal}
  {Advances in colloid and interface science}\ }\textbf {\bibinfo {volume}
  {81}},\ \bibinfo {pages} {167} (\bibinfo {year} {1999})}\BibitemShut
  {NoStop}%
\end{thebibliography}%
\end{document}